\documentclass[seceq]{ptptex}

\usepackage{graphicx}

\newcommand{\be}{\begin{equation}}
\newcommand{\ee}{\end{equation}}
\newcommand{\bea}{\begin{eqnarray}}
\newcommand{\eea}{\end{eqnarray}}
\newcommand{\beann}{\begin{eqnarray*}}
\newcommand{\eeann}{\end{eqnarray*}}
\newcommand{\bma}{\begin{array}{cc}}
\newcommand{\ema}{\end{array}}
\newcommand{\fr}{\frac}
\newcommand{\ra}{\rangle}
\newcommand{\la}{\langle}
\newcommand{\li}{\left}
\newcommand{\re}{\right}

\newcommand{\eps}{\epsilon}

\newcommand{\lam}{\lambda}

\newcommand{\HamO}{\hat{H}_0}
\newcommand{\Ham}{\hat{H}}
\newcommand{\HamV}{\hat{V}_c}
\newcommand{\seteps}{ \{ \eps \} }
\newcommand{\setlam}{ \{ \lam \} }

\newcommand{\Deltaml}{d}
\newcommand{\Deltaov}{\Delta}

\newcommand{\vc}{v_c}

\newcommand{\br}{{\bf r}}

\title{
Many-body effects in the mesoscopic x-ray edge problem
}

\author{
Martina \textsc{Hentschel}${}^1$,
Georg \textsc{R\"oder}${}^1$, and
Denis  \textsc{Ullmo}${}^{2,3}$
}

\inst{
${}^1$ Max Planck Institute for Physics of Complex Systems, N\"othnitzer Str. 38, D-01187 Dresden, Germany\\
${}^{2}$ CNRS, LPTMS UMR 8626, 91405 Orsay Cedex, France \\
${}^{3}$ Univ.\ Paris-Sud, 91405 Orsay Cedex, France \\}

\abst{
Many-body phenomena, a key interest in the investigation of
bulk solid state systems, are studied here in the context of the
x-ray edge problem for mesoscopic systems.  We investigate the
many-body effects associated with
the sudden perturbation following the x-ray excitation of a core electron
into the conduction band. For small systems with
dimensions at the nanoscale we find considerable deviations from the
well-understood metallic case where
Anderson orthogonality catastrophe and the
Mahan-Nozi\`{e}res-DeDominicis response cause characteristic deviations of the
photoabsorption cross section from the naive expectation. Whereas the $K$-edge
is typically rounded in metallic systems, we find a slightly peaked $K$-edge in
generic mesoscopic systems with chaotic-coherent electron dynamics. Thus
the behavior of the photoabsorption cross
section at threshold depends on the system size
and is different for the metallic and the mesoscopic case.
}

\begin{document}

\maketitle

\section{Introduction}
Many-body phenomena such as the Kondo effect or Fermi edge
singularities (FES) have been a key interest in condensed matter physics
for many years\cite{mahan:book}.
Motivated by the experimental progress in the
field of mesoscopic physics and quantum chaos\cite{mesoquantchaos}, especially
the growing interest in many-body effects in those systems\cite{mesomanybody},
we report here theoretical results on the mesoscopic x-ray edge problem.
We are in particular
interested in phenomena associated with a sudden perturbation of
a mesoscopic system
such as a quantum dot or a metallic nanoparticle. We predict
substantial differences to the
metallic case that are falsifiable in state-of-the-art experiments.

In the x-ray edge problem\cite{ohtakaRMP}, a
sudden, localized perturbation is caused by an x-ray exciting a
core electron into the conduction band, leaving a core hole
behind. The response of the conduction electrons to the resulting
attractive potential leads to Anderson orthogonality catastrophe
(AOC)\cite{andersonAOC} -- the overlap between the unperturbed and perturbed
many-body wavefunctions vanishes in the thermodynamic limit. AOC
competes with a second many-body effect
known as Mahan's exciton\cite{mahan:book}
or Mahan-Nozi\`{e}res-DeDominicis\cite{nozieres} (MND) response.
In the metallic case, this leads to Fermi edge singularities, i.e.,
deviations from the naively expected photoabsorption cross section
in the form of a peaked or rounded edge. More precisely, the behaviour
at threshold is known to follow a power law, with the exponent 
determined by the partial wave phase shifts $\delta_l$ at the Fermi energy
in response to the sudden perturbation ${\hat V}_c$ for orbital
channel $l$ \cite{ohtakaRMP} ($\omega$ is the x-ray energy, $\omega_{\rm th}$
indicates the threshold energy),
\begin{equation}
A(\omega) \propto (\omega-\omega_{\rm th})^{-2 \frac{|\delta_{l_0}|}{\pi}
+ \sum_l 2 (2 l+1) \left[ \frac{\delta_l}{\pi} \right]^2} \:.
\label{Aofw_thres}
\end{equation}
The two terms in the exponent have opposite signs and correspond to the MND
response (with $l_0$ being the optically excited channel)
and to AOC, respectively.

The many-body enhancement depends, via the dipole
selection rules, on the symmetry relation between the core and local conduction
electron wavefunction. The MND response will be non-vanishing only if the dipole
selection rules are fulfilled.
Assuming the local part of the conduction electron wavefunction
to be of $s$-type, we distinguish
between core electrons with $s$-symmetry ($K$-shell, $l_0=1$) and $p$-symmetry
($L_{2,3}$-shell, $l_0=0$)\cite{citrin} and refer to the photoabsorption threshold
as $K$- or $L$-edge, respectively. In metals, the phase shifts are such that the
$K$-edge is typically rounded whereas the $L$-edge is peaked.
In the following, we will apply the usual model
of a spherically symmetric potential ${\hat V}_c$ \cite{ohtakaRMP} such that
$\delta_l=0$ for $l>0$. The origin of the form of the FES typically
observed in metals becomes then immediately apparent\cite{ohtakaRMP, citrin}.

For coherent systems
with chaotic or regular dynamics
we use a random matrix model \cite{RMT} or exact solutions of the
Schr\"odinger equation, respectively, to compute the AOC overlap
and the photoabsorption
cross section. For the latter we use the Fermi golden rule approach
introduced by Tanabe and Ohtaka \cite{ohtakaRMP}. Our model applies
to nanoparticles and quantum dots with chaotic or regular (for
example circular) shape, respectively. Comparing our results with
the well-understood metallic problem, we find substantial
changes\cite{fesprl,fesprb}: (1) the finite number of particles
causes AOC to be incomplete, (2) the sample-to-sample fluctuations
of the discrete energy levels produce a distribution of AOC
overlaps, and (3) most importantly, the dipole matrix elements
connecting the core and conduction electrons are substantially
modified. One of our key results is that a photoabsorption cross
section showing a rounded edge in metals will change into a slightly
peaked edge on average as the size of a chaotic system is reduced
to the mesoscopic-coherent scale. This peak is a direct signature of
a coherent-chaotic dynamics of the conduction electrons reached in
the mesoscopic regime: It is this property that leads to a
non-vanishing dipole matrix element\cite{fesprl} and therefore to a
situation that is reminescent of the $L$-edge behaviour. We will
come back to this in more detail below.

The outline of the paper is as follows. First, we consider AOC for
mesoscopic systems subject to a sudden, rank-one perturbation. Then,
we present results on the photoabsorption cross section in those systems.
Our conclusion includes a discussion of possible experimental
setups that allow for a verification of our prediction.

\section{Anderson orthogonality catastrophe in mesoscopic systems}

We describe the non-interacting system by a
Hamiltonian $ \HamO = \sum_{i , \sigma} \eps_i c^\dagger_{i,\sigma}
c_{i,\sigma}$, where $c^\dagger_{i,\sigma}$ creates a particle with spin $\sigma
= \pm $ in the orbital $\varphi_i(\br)$ ($i=0,\ldots,N-1$).
The unperturbed energy levels $\eps_i$ are system specific and provide
together with the  $\varphi_i$ a unique characterisation of the system.
As reference point, we define the {\it bulklike} system where the
energy levels $\{\eps_i\}$ are spaced equidistantly and the wavefunctions are
constant. 

We furthermore assume that the perturbing potential
is a contact potential $\HamV = {\cal V} \vc
|\br_c \ra \la \br_c |$, with $\br_c$ the location of the core hole and ${\cal V}$
the volume in which the electrons are confined. The diagonal form of the
perturbed Hamiltonian is $\Ham = \HamO + \HamV = \sum_{i , \sigma} \lambda_i
\tilde c^\dagger_{i,\sigma} \tilde c_{i,\sigma}$, where $\tilde
c^\dagger_{i,\sigma}$ creates a particle in the perturbed orbital $\psi_i(\br)$.
For relations between the $\{\eps_i\}, \{\lambda_i\}$ and
$\{\varphi_i\}, \{\psi_i\}$ we refer the reader
to Refs.\cite{ohtakaRMP,fesprb}.

A  remarkable  property of a rank-one perturbation such as a contact potential
is that  all  the quantities of interest  for the  x-ray  edge problem  can be
expressed  in  terms of  the $\seteps$  and $\setlam$ (or otherwise, for example
the wavefunction derivative needed for the dipole matrix element,
can be taken as independent random variables following a known,
often a Porter-Thomas, distribution).
Ignoring for now the spin variable,
the  overlap between the  many body ground states  
with $M$ particles of $\HamO$ and $\Ham$, $|\Phi_0 \ra $
and $|\Psi_0 \ra$, can be expressed as \cite{ohtakaRMP}
\begin{equation}
\label{eq:overlap}
|\Deltaov|^2=
|\la \Psi_0 | \Phi_0 \ra |^2 = \prod_{i=0}^{M-1}  \prod_{j=M}^{N-1}
\fr{ (\lam_j - \eps_i)  (\eps_j - \lam_i) }{ (\lam_j - \lam_i)  (\eps_j - \eps_i)} \:.
\end{equation}

For the Fermi energy in the middle of the conduction band,
the phase shift $\delta_0$, the perturbation strength $\vc$ and
the mean level spacing $\Deltaml$ are related through
$\delta_0= \arctan (\pi {\vc}/{\Deltaml})$ \cite{ohtakaRMP}
($\delta_0$ is negative since the core potential is attractive).
In our case, it turns out to be necessary to
take boundary effects into account \cite{fesprb}, which, in addition to the formation
of a bound state that is discussed in more detail in Ref.\cite{fesprb},
modify the phase shifts
away from the band center. This can be included
simply by introducing a variable $v_{i}$
given by
\begin{equation}
\fr{1}{v_i}=\fr{1}{\vc} + \fr{1}{\Deltaml} \ln \fr{N+0.5-i}{i+0.5}
                        \:.
                        \nonumber
\end{equation}
for $i \in (1,N/2)$, and the analogous form for $i \in (N/2,N)$. This gives
rise to level-dependent phase shifts $\delta_i$.
It is known from the metallic case that the phase shift at the Fermi energy
determines the FES.
We will now address the question
to what extent this statement holds in the mesoscopic
case. 

\begin{figure}[t]
  \begin{center}
  \includegraphics[width=12cm]{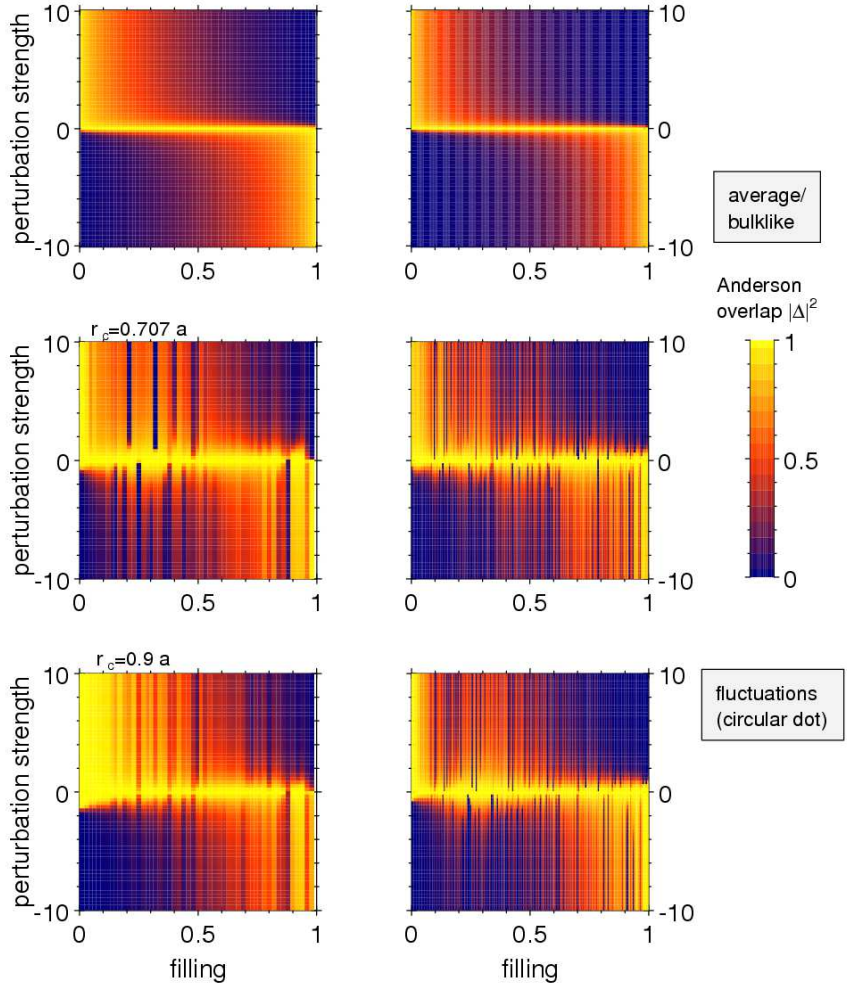}
  \end{center}
  \caption{Anderson orthogonality catastrophe in bulklike systems (top row) and individual
  mesoscopic systems of different sizes. The larger (smaller)
  system with space for up to 1000 (100) electrons in the
  conduction band is shown on the right (left). The mesoscopic system considered here
  is a circular quantum dot of radius $a$. The radial position $r_c$ of the
  perturbation is slightly different in the central and lower row, respectively, giving
  rise to different behavior of the individual systems.}
\label{fig1}
\end{figure}

In Fig.~\ref{fig1}, we first
discuss AOC for two mesoscopic systems of different size
(with space for $N$=100 and $N$=1000 electrons in the
conduction band, cf.~left and
right panels, respectively).
The AOC overlap as a function of filling
of the conduction band
and the perturbation strength is shown in color scale
(for reasons of symmetry, we also consider
positive $v_c$ in addition to the attractive $v_c<0$ describing the effect of
the core hole). In the top row, the situation in the bulklike case is shown.
Clearly, for otherwise equal parameters, the overlap becomes
smaller for larger systems (i.e. closer to the thermodynamic limit).
There is, however, a considerable amount of structure visible beyond this. The
somewhat counterintuitive increase of the overlap with increasing filling is a
property of the rank-one model that we use, in particular of the 
level-dependent phase sifts discussed above. More precisely, within this model the
phase shift is, for $v_c<0$,
larger for smaller fillings, see Ref.~\cite{fesprb} for details.
In other words, the experimentally relevant 
phase shift at the Fermi energy depends on both the filling and
the perturbation strength $v_c$, and one and the same phase shift can be realized
using different sets of parameters, cf.~the discussion in the context of
Fig.~\ref{fig2} below. 

\begin{figure}
  \begin{center}
  \includegraphics[width=14cm]{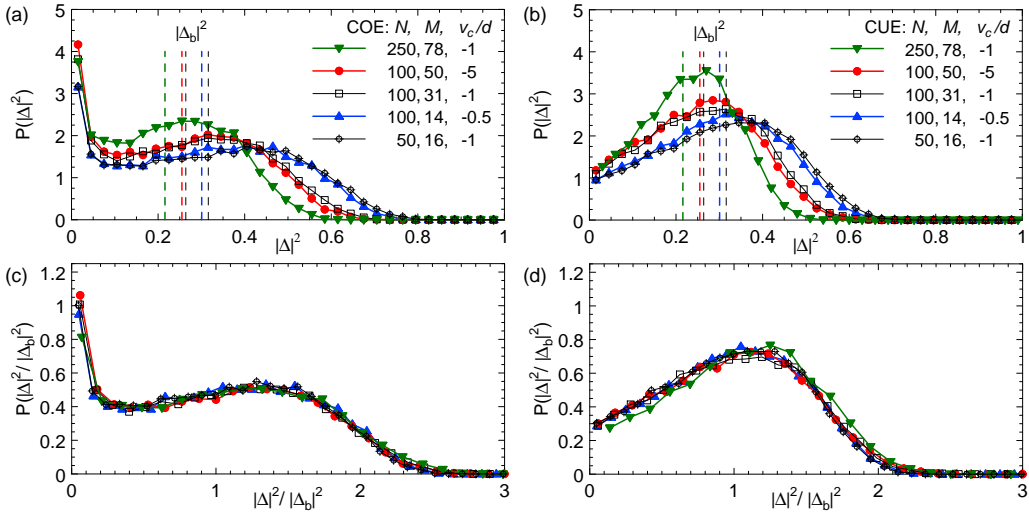}
  \end{center}
  \caption{Distribution of the Anderson overlap $|\Delta|^2$ in chaotic mesoscopic systems.
  The same phase shift $\delta_F \approx -\pi/2$
  at the Fermi energy is realized using different parameter
  sets $\{N,M,v_c/d\}$ for random matrices from (a) the COE and (b) the CUE ensemble.
  Once the values $|\Delta|^2$ are scaled by the system-size dependent bulklike
  overlap $|\Delta_b|^2$, the distributions all coincide: The fluctuations of the overlap
  depend only on the phase shift $\delta_F$ at the Fermi energy.}
  \label{fig2}
\end{figure}

The central and lower row of Fig.~\ref{fig1} show
snapshots of some corresponding
mesoscopic cases. More specifically, a regular quantum dot of disk shape
and with hard walls
is considered. To this end, the Schr\"odinger equation is solved
exactly for the energy levels and wavefunctions.
The perturbation is placed at two different
locations $\br_c$. Fluctuations characteristic for mesoscopic systems
are clearly visible.
In particular,
the overlap is not any more a monotonous function of filling:
Changing the filling corresponds to changing the orbital at
the Fermi energy. More important than a (model-specific)
trend in the phase shift
is now the distribution of energy levels
around the Fermi energy and the amplitude of the
wavefunctions at the position of the perturbation. This also explains the
sensitivity of the Anderson overlap
against changes of the location 
of the perturbing potential that are clearly visible when comparing the
central and lower row of Fig.~\ref{fig1}.

In order to gain a better understanding of the big mesoscopic
fluctuations, 
we will now turn to the universal case
of {\it chaotic} mesoscopic systems,
regular systems such a circular quantum dots
will be considered elsewhere\cite{fesregular}.
We use a random matrix model to effectively describe
the (non-interacting) conduction electrons
in the absence of the core hole, i.e., we assume
the unperturbed energy levels $\{\eps_i\}$ to be the eigenvalues of a random
matrix (belonging to the circular orthogonal or unitary ensemble\cite{RMT},
COE or CUE, respectively).
The single particle wavefunctions will then show the characteristic Porter-Thomas
probability distributions \cite{PT}
characterizing the spatial dependence of the wavefunction intensity.
Ensembles of 10000 individual chaotic systems are generated in a Metropolis
algorithm\cite{fesprb}; the joint probability distribution for
the $\{\eps_i\}$ and $\{\lam_i\}$,
the basic ingredient for this method, 
was derived in Ref.\cite{kostya}. The Anderson overlap
(and later on the photoabsorption cross section) is computed for each
realization from Eq.~(\ref{eq:overlap}). Subsequently, average values,
probability distributions, etc. are easily determined.

That even in mesoscopic systems the phase shift
at the Fermi energy is 
a physically important quantity
as known from the metallic
case, becomes clear in Fig.~\ref{fig2}.
In Fig.~\ref{fig2}(a,b), probability distributions of the
Anderson overlap are compared for mesoscopic systems
of various sizes possessing
time reversal symmetry (COE case, on the left) or not
(CUE case on the right). 
All parameter combinations $\{N,M,v_c/d\}$ 
yield a very similar
phase shift $\delta_F \sim -\pi/2$ at the Fermi energy.
Nonetheless, i.e., unlike the expectation based on
the behavior of bulk systems would suggest,
the probability distributions are rather
different. But so are the reference bulklike values for the Anderson overlap
indicated by the dashed lines
(assignment is such that $\Delta_b$ increases
in the order of the legend entries). Indeed, the probability
distributions for the overlap scaled by the corresponding bulklike value
convincingly coincide [lower panels (c) and (d)]:
In the mesoscopic case, the overlap does depend on the system
size $N$ but the mesoscopic fluctuations are solely determined by the value of the
phase shift.

Eventually, we point out the finite probability for finding
zero overlap, that moreover is rather different for the COE and CUE case,
respectively. Referring the interested reader to the details given
in Ref.\cite{fesprb}, the distinctive behavior traces back to the differences
of the Porter-Thomas distribution for finding small values 
in the presence or absence
of time-reversal symmetry, respectively.

\section{Photoabsorption spectra: From rounded to peaked edge}

Next, we discuss the absorption spectra, thereby focussing especially on the
$K$-edge. Our approach is based on Fermi's golden rule following the work
by K.~Ohtaka and Y.~Tanabe \cite{ohtakaRMP} who showed that this method provides
a comprehensive description of the x-ray edge problem.
The photoabsorption cross section in the mesoscopic case is then obtained from
(using units $\hbar=1$) 
\begin{equation}
  A(\omega) =2 \pi 
  \sum_f | \la \Psi_f | \hat D | \Phi_0^c \ra |^2
  \delta(E_f - E_0^c - \omega) \:,
\label{Aofw_Fermigoru}
\end{equation}
where the sum is taken over all perturbed final states $\Psi_f$ (of energy $E_f$)
connected  to the
unperturbed groundstate $\Phi_0^c = \prod_{\sigma=\pm } \prod_{j=0}^{M-1}
c^\dagger_{j,\sigma} c^\dagger_c |0\ra$ 
(of energy $E_0^c$)
by the dipole operator $\hat D$ 
($c^\dagger_c$ creates the core electron 
in the empty band $|0\ra $).
We are interested in processes involving the core hole; thus, the
dipole operator can be written as $\hat D = \mbox{const.  } \sum_{j=0}^N \li(
w_{j} \tilde{c}_{j \sigma}^\dagger {c}_c + h.c.  \re)$. At $K$-edge, the core
electron wavefunction and the local part of the conduction electron wavefunction
are both of $s$-symmetry, $w_j$ is related
to the {\it derivative} of the perturbed orbital
$\psi_j$ in the direction $\vec e$ of the polarization of the x-ray through $w_j
= \vec{e} \cdot {\bf \nabla} \psi_j (\br_c)$.

We first turn to the absorption cross section right at threshold,
$\omega= \omega_{\rm th}$, and neglect the spin degree of freedom for the
moment\cite{mahan:book, nozieres}.
The only possible final state is then $\Psi_{f^0} =  \prod_{j=0}^{M}
\tilde{c}_{j}^\dagger |0\ra $.
Without a perturbing potential, the only contribution is the direct process $
w_{M}  \tilde{c}_{M}^\dagger  \tilde{c}_c $. In the presence of a
perturbation, however, the new and old orbitals are not identical.
This implies that the so-called {\it replacement processes}, terms
with $j < M$, also contribute coherently, giving
\cite{ohtakaRMP}
\begin{equation}
 |\la \Psi_{f^0} | \hat D | \Phi_0^c \ra |^2  \propto
 |w_{M} \Deltaov|^2   \li| 1 - \sum_{i=0}^{M-1}
                           \fr{ w_{i } \Deltaov_{\bar i, M}}
                              { w_{M} \Deltaov }  \re | ^2
\label{dirrepl}
\end{equation}
where $\Deltaov_{\bar i, M}$ is defined by generalizing Eq.~(\ref{eq:overlap})
with level $i \, (<M)$ replaced by $M$.  Since for chaotic systems the
derivative of the wavefunction, $ {\cal V} k^{-2} \times | \nabla_{\vec{e}}
\psi_j|^2$, is known to have Porter-Thomas fluctuations uncorrelated with the
wavefunction itself \cite{prigodin}, we can proceed as for the overlap to
construct the distribution of $|\la \Psi_{f^0} | \hat D | \Phi_0^c \ra |^2 $.

Away from threshold, part of the x-ray energy can excite additional electrons
above the Fermi energy in so-called {\it shake-up processes}.  Their contribution is a
straightforward generalization of Eq.~(\ref{dirrepl}).  Although the number of these
processes grows in principle exponentially with the energy of the x-ray, only few
shake-up processes contribute significantly to the photoabsorption.
Shake-up processes involving more than three shake-up pairs can safely be
neglected\cite{we_FESII} as was also found previously\cite{few_shakeup_suffice}.
The spin of the electrons is taken into account by including the AOC contribution
due to the additional electronic channel.

\begin{figure}
\begin{center}
  \includegraphics[width=10cm]{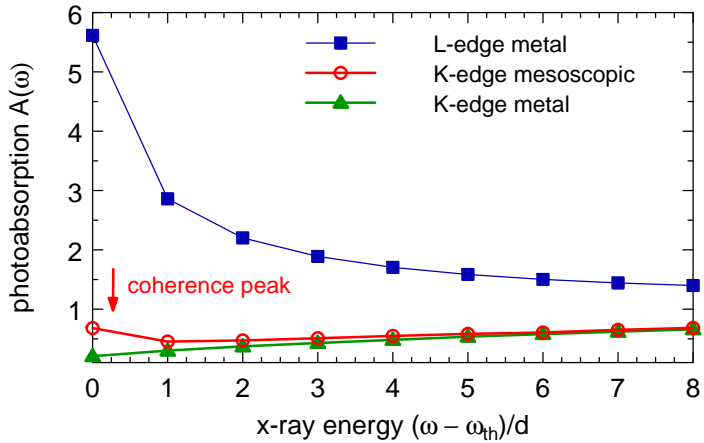}
\end{center}
  \caption{Photoabsorption spectra for chaotic-coherent mesoscopic systems
  and bulklike metals in comparison. Whereas the $L$-edge is peaked in both cases,
  the typically rounded metallic $K$-edge evolves into a peaked signature when
  the system size is reduced to the chaotic-coherent scale.}\label{fig3}
\end{figure}

The result for the photoabsorption cross section 
at the $K$-edge of a mesoscopic system
is shown in Fig.~\ref{fig3} (open circles). For comparison, the bulklike curves at
$K$- and $L$-edge (filled triangles and squares, respectively) are also provided.
They are obtained assuming equidistant energy levels and constant
dipole matrix elements that explicitly depend on the symmetry relation between
the optically active channel and the core electron \cite{ohtakaRMP}.
As discussed above, 
this typically leads to
a {\it rounded $K$-edge} (vanishing dipole matrix elements, only AOC contributes)
and a {\it peaked $L$-edge} (the MND response, being linear in the phase shift,
overcompensates AOC). Remarkably, we find a behavior reminiscent of such
a peaked edge at the mesoscopic $K$-edge that is, on average, slightly peaked.
This striking difference has its origin
in the chaotic-coherent dynamics of the electrons in generic (ballistic) mesoscopic
systems such as quantum dots or metallic nanoparticles.

\section{Conclusions}

The central result of this work is that changes in the dynamics of
electrons in a Fermi sea may imply characteristic changes in
the many-body response of the system.
They occur, e.g., as a result of a systematic reduction
of the system size from the bulklike-metallic to the
mesoscopic-coherent scale.
One possibility to make these changes
visible is through the photoabsorption cross section in response to the sudden
creation of a localized perturbation following the excitation of a core electron.
In particular, a typically rounded $K$-edge should develop into a slightly
peaked edge when the system size is sufficiently reduced to
induce chaotic-coherent dynamics of the electrons.
This signature, marked by the arrow in Fig.~\ref{fig3},
is an effect of the coherent
confinement in the chaotic system where the dipole matrix element at $K$-edge is
determined by the derivative of the wavefunction that, unlike the bulklike case,
is independent from the wavefunction itself. Most importantly, it will take
non-zero values on average, and consequently lead to a signature that is qualitatively
comparable to the metallic $L$-edge behavior.

Although the effect of the transition to a rounded edge
seems to be rather small and requires resolution of the x-ray
energy on the order of the mean level spacing (cf.~Fig.~\ref{fig3}),
such an x-ray absorption experiment using metallic nanoparticles
should become possible
in the near future.
Using nowadays technology, we suggest experiments using arrays of quantum dots.
The excitation would then not occur by an x-ray and from the
core level, but rather by radiation from a microwave laser and from an impurity
level specifically introduced by doping
in between the valence and conduction band: The physics
that we describe here, namely the sudden perturbation of a Fermi sea
of electrons by a localized potential, is the very same.
The available energy resolution and manageability
allow one, in principle, to determine the average values
of the photoabsorption cross section and the signature of a coherence peak
at the $K$-edge threshold.

\section*{Acknowledgments}
We would like to sincerely thank Kazuo Ohtaka and Yukito Tanabe for
illuminating discussions and hospitality at Chiba University.
We thank Harold U. Baranger for many discussions and for attracting our interest
to this topic. We also thank  Boris Altshuler, Swarnali Bandopadhyay,
Yuval Gefen, Igor Lerner, Kostya Matveev, Dima Shepelyanski, Jens Siewert, 
and Igor Smolyarenko
for useful and stimulating discussions.

%

\end{document}